\documentclass[%
 reprint,
showpacs,preprintnumbers,
nofootinbib,
nobibnotes,
 amsmath,amssymb,
 aps,
prb,
floatfix,
]{revtex4-1}

\usepackage{graphicx}
\usepackage{dcolumn}
\usepackage{bm}

\usepackage{soul}

\DeclareGraphicsExtensions{.pdf,.jpg, .png}
\usepackage{epstopdf}
\usepackage{enumitem}

\begin{document}
\preprint{}

\title{Spinning Radiation from Topological Insulator}

\author{Emroz Khan and Evgenii E. Narimanov}
\address{School of Electrical and Computer Engineering and Birck Nanotechnology Center, Purdue University, West Lafayette, Indiana 47906, United States}



%

\date{\today}
\begin{abstract}
We show that thermal radiation from a topological insulator carries a nonzero average spin angular momentum.
\end{abstract}


\pacs{44.40.+a, 03.65.Vf, 73.20.−r}

\maketitle


Thermal radiation, both at far and near field, has applications in many areas ranging from radiative cooling \cite{raman2014passive}, thermophotovoltaic systems \cite{lenert2014nanophotonic} and thermal imaging \cite{gaussorgues2012infrared}. Nanolythography techniques \cite{li2018nanophotonic} can tailor the radiation by engineering structures having lengthscales complarable to the characteristic wavelength of thermal emission. In particular, chiral photonic crystal having a polarization dependent bandgap \cite{lee2007circularly} or metasurface with unit cells lacking mirror symmetry \cite{li2015circularly} have been reported to emit circularly polarized radiation. However owing to the design and fabrication complexity \cite{dyakov2018circularly}, it is desirable to have thermal radiation with circular polarization property from a material having simple planar geometry without any added layered or surface structure. 

A solution to this problem can be offered by topological insulator\cite{hasan2010colloquium} (TI) with its property of mixing electric and magnetic fields \cite{qi2008topological}. The corresponding magnetoelectric coupling can cause an electric field to induce magnetization and a magnetic field to induce polarization in these materials giving rise to unusual electromagnetic phenomena including induction of image magnetic monopole \cite{qi2009inducing}, half-integer quantum Hall effect\cite{bernevig2006quantum}, Faraday rotation \cite{karch2009electric} etc. This new physics also manifests itself in the thermal emission from TIs, as we show in this letter, in the form of emergence of spin angular momentum in the radiation.

Under the equilibrium interaction with the environment, TIs will have energy contribution to the thermal radiation from its surface state electrons through the inelastic backscattering with phonons \cite{giraud2011electron}. Spin-momentum locking of the scattered electrons will result in a nonzero correlation among the fluctuating spin currents. This would lead to a thermal radiation which will have some degree of circular polarization and hence will carry a nonzero average spin angular momentum.

\begin{figure}[tb]
\centering
   \includegraphics[width=1.5 in]{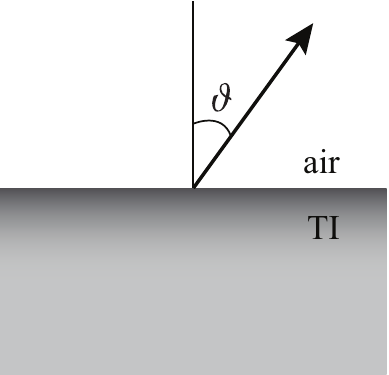}
\caption{\footnotesize Schematic of a semi-infinite slab of a topological insulator (TI) as a thermal emitter into air. }
\label{Fig:schematic}
\end{figure}

 
For design simplicity we consider a semi-infinite slab of a TI, described by permittivity $\epsilon$, with an air interface as shown in Fig. \ref{Fig:schematic} (without loss of generality, the subsequent analysis will inherently restrict our attention to the helical states only at one edge). In the presence of the magnetoelectric coupling, Maxwell equations for TIs retain the usual form\cite{qi2008topological}  with a modified displacement current $\textbf{D}=\epsilon \textbf{E}-\kappa \textbf{B}$ and magnetic field $\textbf{H}= \textbf{B}/\mu+\kappa \textbf{E}$ where $\kappa = \alpha \Theta$ plays the the role of magnetoelectrical susceptibility with $\alpha$ being the fine structure constant and $\Theta$ being the (quantized) axion field\cite{wilczek1987two}. While this axion model \cite{qi2008topological} is limited to the frequency below the bulk bandgap, the corresponding cutoff in actual TIs is well above the thermal frequency range \cite{xia2009observation}, which justifies the use of the model for thermal purposes.

At equilibrium the TI slab at temperature $T$ will act as a thermal bath of photons with the emission flux depending on transmission from TI into the air. From Rytov's formulation of fluctuation electrodynamics \cite{rytov1959theory} the state of polarization of these thermal photons will be given by the stationary eigenstates of the scattering matrix, which in this case is simply the reflection matrix of the slab-air interface. For a usual isotropic dielectric these eigenstates would be the usual linear $|s\rangle$ and $|p\rangle$ polarized light.

Due to the magnetoelectric coupling the eigenstates inside a TI become a linear combination of $|s\rangle$ and $|p\rangle$ polarizations given by
\begin{equation}
|q_{\pm}\rangle = \xi_{\pm} \left( \eta |p\rangle + (\gamma \pm \sqrt{\eta^2+\gamma^2}) |s\rangle \right), 
\label{Eq:q}
\end{equation}
where $\eta = 2\kappa \sqrt{\epsilon},\: \gamma = 1+\kappa^2-\epsilon$ and $\xi_{\pm}$ are normalization constants: $\langle q_+|q_+\rangle=\langle q_-|q_-\rangle=1$. Note that in the absence of loss the coefficients are real and consequently, the eigenstates are still linearly polarized. However, the two polarization directions are now rotated about the propagation direction while maintaining orthogonality with each other, and are no longer either perpendicular to or confined in the plane of incidence as in the cases for the usual $|s\rangle$ or $|p\rangle$ polarizations, respectively. In this case the emitted photon in the air would still retain linear polarization. Since spin angular momentum for linearly polarized light is zero \cite{berry2009optical}, for a lossless TI each individual photon and the thermal radiation as a whole will not carry any spin angular momentum. 

However, presence of a nonzero loss is essential for emission of thermal radiation. Because of the noticeable absorption in existing TIs \cite{esslinger2014tetradymites}, the eigenstates for the thermal photon become complex linear combinations of $|s\rangle$ and $|p\rangle$ polarization. As a result, the eigenstates now show elliptic polarization and each radiated photon carries a nonzero spin angular momentum. 

Using Rytov's theory for the total spin angular momentum flux in the normal direction (with respect to the TI-air interface) we obtain 
\begin{equation}
J_n  = \int d\omega \int d\vartheta \: \rho(\omega) \frac{c}{n} \cos\vartheta' \left(1- \frac{|r_+|^2+|r_-|^2}{2} \right) S_n \:,
\end{equation}
where 
\begin{equation}
\rho(\omega) = \frac{\omega^2 n^3}{\pi^2 c^3} \frac{1}{\text{exp}\left(\hbar \omega /kT\right)-1}
\end{equation}
is the photon density at a particular frequency $\omega$, $n$ is the real part of refractive index of the TI, $\vartheta'$ is the angle of photon incidence corresponding to the emission angle $\vartheta$ (see Fig. \ref{Fig:schematic}), and $S_n$ is the normal component of ensemble averaged per photon spin angular momentum. Also
\begin{equation}
r_{\pm} =\frac{1}{\Delta} \left(\beta \pm \sqrt{\eta^2+\gamma^2} \right)  
\end{equation}
are the reflection coefficients from TI into the air for the two polarization states $|q_{\pm}\rangle$. Here, $\beta = \epsilon/\nu - \nu$ and $\Delta = - \left(1+ \kappa^2+\epsilon+\epsilon/\nu+\nu  \right)$ with $\nu = k_{n, \text{TI}}/k_{n, \text{air}}$, and the subscript $n$ corresponds to the normal component of the wavevector in the respective media.

When a photon in either of the two eigenstates $|q_{\pm}\rangle$ emerges outside the TI, it will assume a corresponding new state $|w_{\pm}\rangle \propto T|q_{\pm}\rangle $ (with the normalization: $\langle w_{\pm}|w_{\pm}\rangle=1$), where $T$ is the transmission matrix from TI into the air, having a representation in the $|p\rangle - |s\rangle$ basis given by
\begin{equation}
T=-\frac{2}{\Delta}
  \left( {\begin{array}{cc}
    (\nu +1) \sqrt{\epsilon} &  \kappa \nu \\
   - \kappa \sqrt{\epsilon} &  (\epsilon + \nu) \\
  \end{array} } \right).
\end{equation}
The spin angular momentum carried by each such photon will be given by 
\begin{equation}
\textbf{S}_{\pm} = 2 \hbar \:\: \text{Im}\left(\langle p|w_{\pm}\rangle^*\langle s|w_{\pm}\rangle  \right) \:  \hat{\textbf{k}}\: , 
\label{Eq:sam}
\end{equation}
where $\hat{\textbf{k}}$ is the unit vector along the propagation direction and $*$ denotes complex conjugation. Detailed calculation shows that $\textbf{S}_+=\textbf{S}_-$ and neither $\textbf{S}_+$ nor $\textbf{S}_-$ depends on the emission angle $\vartheta$. 

\begin{figure*}[tb]
\centering
   \includegraphics[width=6.8 in]{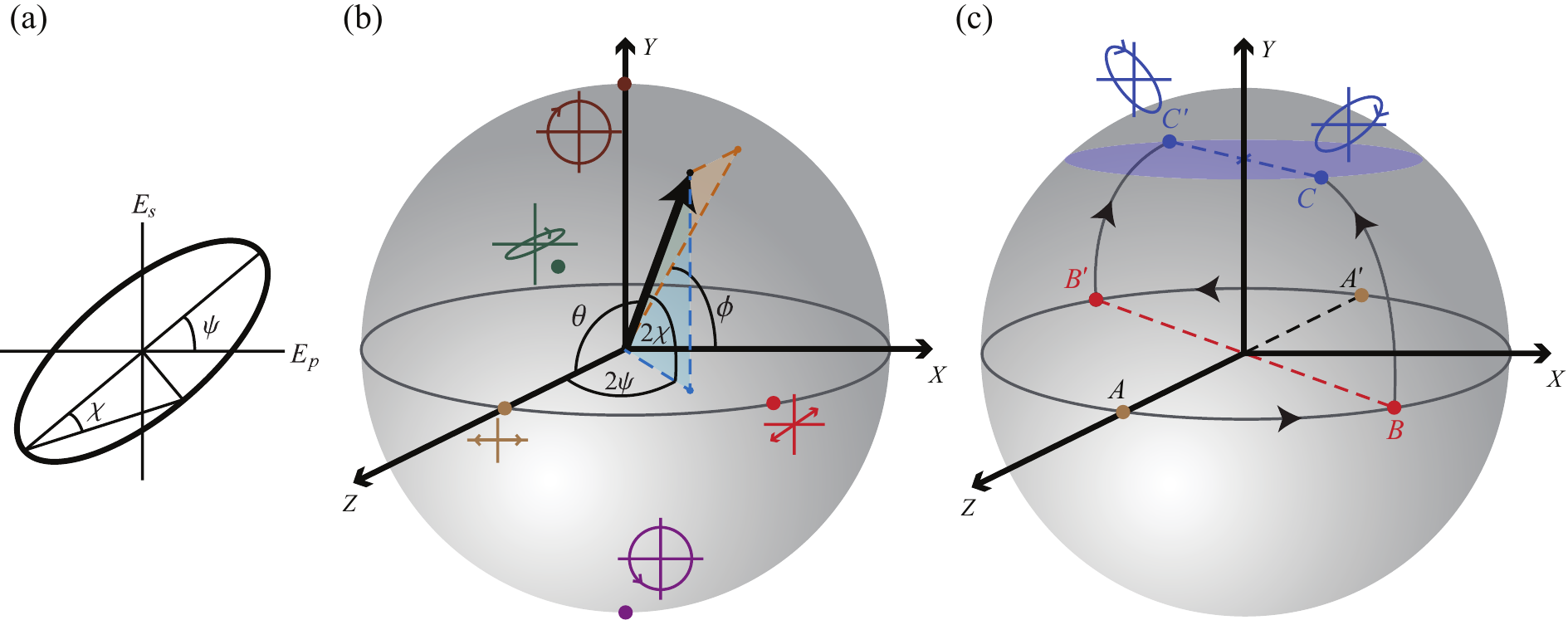}
\caption{\footnotesize Poincar\'e sphere representation for the eigenstates of thermal photons. (a) The polarization state of light having $p$ and $s$ components $E_p  = \cos\frac{\theta}{2}$ and $E_s = e^{i \phi} \sin\frac{\theta}{2}$ can be completely determined by two new angle parameters \cite{born2013principles}: $\psi$ and $\chi$ which give the  tilt angle and a measure of eccentricity of the polarization ellipse, respectively ($0\leq\psi<\pi$ and $|\chi| \leq\pi/4$ with $\chi < 0$ for electric field rotating in a right-hand sense with respect to propagation direction). (b) The corresponding state is represented (by the normalized Stokes vector) as a point on the surface of Poincar\'e sphere: the two poles correspond to circular polarization while points on the equator are linearly polarized. Any other point between these two extrema would denote an elliptic polarization with continuous variation in tilt and eccentricity. Note that the stokes parameter given by the $Y$ coordinate $y =\sin 2\chi  = \sin \theta \sin\phi$ directly gives the spin angular momentum of the photon in units of $\hbar$. (c) The two eigen-polarization states for thermal photons coming from a usual (lossless or lossy) dielectric (points $A$ and $A'$) are the usual $p$ and $s$ polarized light. Introduction of magnetoelectric coupling in the absence of loss  shifts the two eigenstates along the equator (points $B$ and $B'$, the dashed line represents their being diametrically opposite) but still the corresponding photons do not carry any spin angular momentum. Finally, addition of loss lifts the points off the equator (points $C$ and $C'$) and the thermal photons from lossy TI start carrying some degree of circular polarization.}
\label{Fig:stokes}
\end{figure*}

To visualize these results, we represent the polarization state of a photon described by $\cos\frac{\theta}{2} |p\rangle +  e^{i \phi} \sin\frac{\theta}{2}\: |s\rangle$ as a point on the Poincar\'e sphere (as shown in Fig. \ref{Fig:stokes}) where the three axes $X, Y$ and $Z$ denote the three Stoke's parameters. Since $\textbf{S} = \hbar\sin2\chi \:\hat{\textbf{k}}$  (see Eq. (\ref{Eq:sam}) and Figs. \ref{Fig:stokes}(a)-\ref{Fig:stokes}(b)), the $Y$ coordinate directly gives the spin angular momentum per photon in units of $\hbar$ along the propagation direction. 

We also note that the reflection matrix from TI to air can be represented in the $|p\rangle - |s\rangle$ basis as
\begin{equation}
R = \frac{1}{•\Delta}\left(\beta\, \textbf{I} + \pmb{\sigma}\cdot\textbf{B} \right),
\label{Eq:bloch}
\end{equation}
where $\textbf{B}$ is a constant vector $\textbf{B}=( \eta,0,-\gamma)$, whose components are related to the parameters of our topological system, and $\pmb \sigma$ is a vector containing Pauli matrices as its components \cite{blundell2003magnetism}. The functional form of Eq. (\ref{Eq:bloch}) then allows us to interpret the eigenstates of $R$ as those of an effective Hamiltonian matrix $\mathcal{H} = \pmb{\sigma}\cdot\textbf{B}$ for an electron in a magnetic field $\textbf{B}$.

Now for a usual dielectric the effective magnetic field becomes  $\textbf{B} = ( 0,0,1-\epsilon)$ and the resulting eigenstates are the two points on the $Z$ axis corresponding to the $|p\rangle$ and $|s\rangle$ polarization (points $A$ and $A'$ in Fig. \ref{Fig:stokes}(c)). Introduction of loss or variation of the incidence angle does not alter these two polarization coordinates. 

For a lossless TI the effective field  $\textbf{B}$ is real and the eigenstates are represented by a rotated pair of antipodal points on the equator along the support of $\textbf{B}$  (points $B$ and $B'$ in Fig. \ref{Fig:stokes}(c)). These corresponds to photons inside TI having polarization states $|q_{\pm}\rangle$ given by a linear combination of $|p\rangle$ and $|s\rangle$ with real coefficients, i.e., linearly polarized states.\cite{chang2009optical} Since the transmission matrix $T$ is also real in this case, the states of the emitted photon $|w_{\pm}\rangle$ still lie on the equator carrying no spin angular momentum.

However, for a lossy TI the effective field $\textbf{B}$ becomes complex and the corresponding eigenstates are no longer confined in the equator. Since $\mathcal{H}$ is no longer Hermitian, the two eigenstates instead of being antipodal pairs ($\langle q_- | q_+ \rangle \neq 0$) become diametrically opposite points having the same $Y$ coordinates (points $C$ and $C'$ in Fig. \ref{Fig:stokes}(c)) as they maintain an orthogonality relation without complex conjugation $q_-^T q_+ = 0$.\footnote{As the symmetry $\mathcal{H}^T = \mathcal{H}$ remains in effect even for a lossy TI.} Since $\mathcal{H}$ does not contain any angle parameter, polarization of the eigenstates do not depend on the incidence direction of photons. While the transmission matrix $T$ explicitly depends on the incidence angle $\vartheta'$,  for any value of $\vartheta'$ it maps the eigenstates of $R$ (see Eq. (\ref{Eq:q})) to fixed states $|w_{\pm}\rangle$ in the Poincar\'e sphere.\footnote{owing to the fact that $T$ and $R$ are not independent of each other.} Moreover, the eigenstates of the emitted photons also obey $w_-^T w_+ = 0$. Therefore each radiated photon carries an equal amount of nonzero spin angular momentum for both eigen-polarizations, in all emission directions, with the normal component given by
\begin{equation}
S_n =  \frac{4\hbar\, \kappa\, {\rm Im}[\zeta] }{4  \kappa^2 + |\zeta - 2 \kappa^2|^2}  \:   \cos \vartheta,
\label{Eq:Sn}
\end{equation}
where $\zeta = \gamma + \sqrt{\eta^2+\gamma^2}$.

\begin{figure*}[tb]
{\centering
   \includegraphics[width=6.3 in]{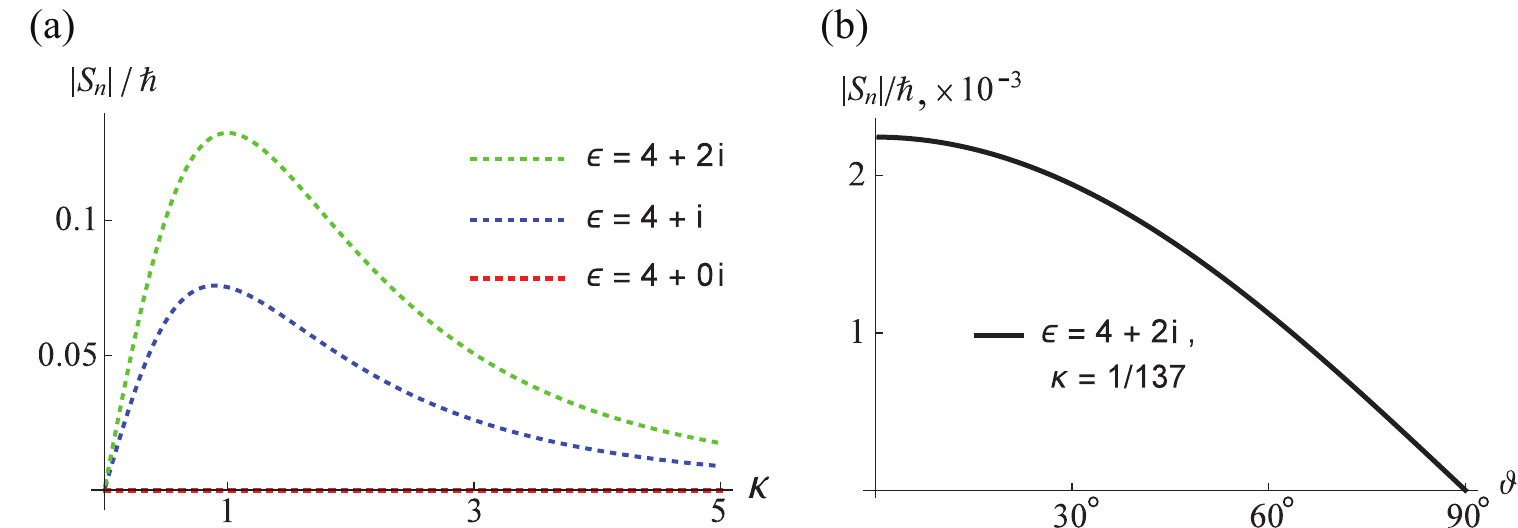}}
\caption{\footnotesize Spin angular momentum in thermal radiation from a lossy TI. (a) Magnitude of the normal component of spin angular momentum per photon emitted at a certain angle, for different values of magnetoelectric coupling $\kappa$ and loss in the medium. Note that the spin angular momentum is nonzero when both coupling and loss are present. Dotted lines have been used to show its functional dependence on the coupling constant as $\kappa$ can only assume values which are integer multiples of the fine structure constant due to quantization of the axion field. (b) Dependence of the normal component of spin angular momentum on the emission angle $\vartheta$.}
\label{Fig:res}
\end{figure*}

Note that both magnetoelectric coupling and material absorption are required for the presence of spin angular momentum in the thermal radiation. This is illustrated in Fig. \ref{Fig:res}(a) which shows the magnitude of $S_n$ (in units of $\hbar$) for a photon emitted at a certain angle as a function of the coupling constant $\kappa$, for different amount of loss in the TI. 

As follows from Eq. (\ref{Eq:Sn}), $S_n \propto \cos \vartheta$; this behavior is illustrated in Fig. \ref{Fig:res}(b) for the typical value of $\kappa = \alpha$.  Note that even though the spin angular momentum per photon along the propagation direction is the same for all emission angles, the normal component varies but does not change sign.

Although the fact that the emitted photons in the two eigen-polarization states carry elliptic polarization (see insets of Fig. \ref{Fig:stokes}(c) near points $C$ and $C'$) not only in the same amount but also in the same rotational sense as well (clockwise or counter-clockwise) may seem counter-intuitive, it does not violate any symmetry of the system, as the spin-momentum locking of surface state electrons in TIs inherently maintains a preferred chirality towards the surface normal\cite{hasan2010colloquium}. Moreover, the emission of spin angular momentum from a TI surface does not break any conservation law in as much the same way as linear momentum is conserved during generation of radiation pressure in thermal emission. 

While the above analysis considered only one side of a TI slab, it can also be applied to a finite slab with thickness greater than the absorption length. The two interfaces in that case will generate spin angular momentum flux in the opposite directions. 

Note that, in contrast to thermal emitters with a structured interface \cite{dyakov2018circularly} that do not radiate light with the same degree of circular polarization in all directions, a slab of TI would radiate an equal amount of spin angular momentum for all emission angles. Offering the control over the spin angular momentum of the thermal radiation, TIs can therefore find potential applications in the generation of structured light from thermal sources.

Furthermore, with the mapping to the effective Hamiltonian of electron in magnetic field via Eq. (\ref{Eq:bloch}), the lossy TI system considered in our present work can be used to realize complex solid state quantum gates, for example the square root of NOT gate\cite{deutsch2000machines}
\begin{equation}
\sqrt{\rm NOT} = \frac{1}{2}
  \left( {\begin{array}{cc}
    1+i &  1-i \\
   1-i &  1+i \\
  \end{array} } \right)
\end{equation}
can be implemented with a low-loss TI (with ${\rm Im}[\epsilon] =  \mathcal{O}(\kappa)$). 

In conclusion, thermal radiation from topological insulator carries nonzero spin angular momentum. This can be of practical importance for generating thermal fields with elliptic polarization, and as a manifestation of the presence of an axion field in a thermal emitter that can be used for its detection and measurement.

The authors acknowledge support for this work from the National Science Foundation, Grant No.  DMREF- 1629276 and the Gordon and Betty Moore Foundation.

\bibliography{refs}

\end{document}